\documentclass[12pt,a4paper]{article}
\usepackage{amsmath}
\usepackage{graphicx}
\begin{document}
\begin{titlepage}
\title{ The LHC data and an upper bound for the inelastic diffraction.}
\author{ S.M. Troshin, N.E. Tyurin\\[1ex]
\small  \it Institute for High Energy Physics,\\
\small  \it Protvino, Moscow Region, 142281, Russia}
\normalsize
\date{}
\maketitle

\begin{abstract}
We  comment on the status of the Pumplin bound for the inelastic diffraction in the light of
the recent LHC data for elastic scattering. 
\end{abstract}
\end{titlepage}
\setcounter{page}{2}

The experiments performed at the LHC have confirmed  
continuous increase of the total, elastic and inelastic cross--sections with energy, which was observed at lower energies. 
Those experiments brought us  closer to clarification of the elusive asymptotic regime of 
strong interactions.
Arguments based on analiticity and unitarity of the scattering matrix 
lead to conclusion that the Froissart-Martin bound \cite{froi,martin} for the total cross-sections 
would be saturated at asymptotics.
Indeed, the functional energy dependence of the  total cross-sections is often taken to follow $\ln^2 s$-dependence at very
high energies, but 
the value of the factor in front of $\ln^2 s$ remains  an issue. 
This is 
related to the choice  of the upper limit for the partial amplitude. Namely,  this limit may correspond to the
maximum of the inelastic channel contribution to the elastic unitarity, when
\begin{equation}\label{bd}
  \sigma_{el}(s)/\sigma_{tot}(s)\to 1/2,
\end{equation}
or it corresponds to a maximal value of the partial amplitude allowed by unitarity resulting in the asymptotical limit
\begin{equation}\label{rd}
  \sigma_{el}(s)/\sigma_{tot}(s)\to 1.
\end{equation}
The first option is to be an equivalent of the presupposed absorptive nature of the scattering, while the second option
assumes the alternative which was interpreted as a reflective scattering \cite{reflect}.
With assumption of  the absorptive scattering the original Froissart-Martin bound for the total cross-sections 
has been improved \cite{mart} and the upper bound for the total inelastic cross-section reduced by factor of 4 has also been derived \cite{mart}.

The assumption on absorptive scattering was also crucial under  derivation of the Pumplin bound \cite{pumplin}
for the inelastic diffraction:
\begin{equation}\label{pump}
 \sigma_{diff}(s,b)\leq \frac{1}{2}\sigma_{tot}(s,b)-\sigma_{el}(s,b),
\end{equation}
where  \[\sigma_{diff}(s,b)\equiv \frac{1}{4\pi}\frac{d\sigma_{diff}}{db^2}\] is the total cross--section of  all the inelastic diffractive processes in the impact parameter 
representation and \[ \sigma_{tot}(s,b)\equiv \frac{1}{4\pi}\frac{d\sigma_{tot}}{db^2}\,\, ,\,\,  \sigma_{el}(s,b)\equiv \frac{1}{4\pi}\frac{d\sigma_{el}}{db^2}.\]
The inequality Eq. (\ref{pump}) was obtained in the framework of the Good--Walker formalism for the inelastic diffraction \cite{gwalk}.
Eq. (\ref{pump}) is to be valid for  each value of the impact parameter of the collision $b$. It can be integrated over impact parameter
with the result
\begin{equation}\label{pumpint}
 \sigma_{diff}(s)\leq \frac{1}{2}\sigma_{tot}(s)-\sigma_{el}(s).
\end{equation}
Thus, in the framework of the absorptive scattering approach, the Eqs. (\ref{bd}) and (\ref{pumpint}) should be fulfilled
simultaneously if the black disk limit is supposed to be reached asymptotically, i.e. 
\begin{equation}\label{bdin}
  \sigma_{inel}(s)/\sigma_{tot}(s)\to 1/2
\end{equation}
while
\begin{equation}\label{bdind}
  \sigma_{diff}(s)/\sigma_{tot}(s)\to 0
\end{equation}
and
\begin{equation}\label{bdindi}
  \sigma_{diff}(s)/\sigma_{inel}(s)\to 0
\end{equation}
at $s \to \infty$.
Those  limits are the divergent ones. Indeed,  $\sigma_{diff}(s)$\footnote{A common opinion associates any type of inelastic diffraction with one or several  Pomeron exchanges.} is, by definition\footnote{Cf. for discussion \cite{pred}.}, a leading part of the inelastic
cross--section $\sigma_{inel}(s)$. The experimental data obtained at the LHC demonstrates approximate
energy--independent ratio
$ \sigma_{diff}(s)/\sigma_{inel}(s)$   \cite{alice}.
In contrast to the definion of the inelastic diffraction and available experimental data, one should conclude then, that the inelastic diffraction is, in fact,  a subleading mechanism 
 of the increase of the inelastic cross-section and the main role in this growth 
 belongs to  nondiffractive inelastic processes. 
 Such a statement is not easy to adopt.

There is no such apparent embarassment  in the approach which  suppose saturation of the unitarity limit.
The assumption that unitarity limit is to be saturated asymptotically  leads to a relatively slower increase of the inelastic cross-section
\begin{equation}\label{bdind0}
  \sigma_{inel}(s)/\sigma_{tot}(s)\to 0
\end{equation}
 which allows one to keep considering inelastic diffraction as a leading mechanism  of the inelastic cross--sections growth.
In this approach the ratio  of the elastic to total cross-section (\ref{rd}) corresponds to  energy increase of the total inelastic cross-section slower than  $\ln^2 s$ while Eqs. (\ref{rd}) and  (\ref{bdind0}) take place. It should be noted that available experimental data are consistent with decreasing
dependence of the ratio $ \sigma_{inel}(s)/\sigma_{tot}(s)$ with energy.

The possibility of the black disk limit crossing   was discussed in the general framework of the rational 
unitarization on the base of the CDF data obtained at Tevatron \cite{phl}. It should be noted that the value of $\mbox{Im} f(s,b=0)$ 
has increased from $0.36$  (CERN ISR) to $0.492\pm 0.008$ (Tevatron)  and it is on the edge of the black disk limit in this energy domain\cite{girom}. As it was mentioned in \cite{phl}, the exceeding of the black disk limit turns the Pumlin bound to be groundless.
But, this conclusion deserves to be more specified now. In fact, the Pumplin bound does not valid only in the limited range of the small and moderate values of the collision impact parameter where absorptive approach is not applicable. 
We discuss this point here, but we should mention first that the Pumplin bound
has been obtained with an assumption of the pure imaginary amplitudes of elastic and diffractive scattering. We  use this simplification
here.

The model-independent  reconstruction of the impact--parameter dependent quantities from this experimental data set demonstrates that the black disk limit has been crossed in elastic scattering at small values of $b$ \cite{alkin}. In fact, the elastic scattering $S$-matrix 
element $S(s,b)\equiv 1-2f(s,b)$, where $f(s,b)$ is an imaginary part of the elastic amplitude, is negative at $0<b<0.2$ fm and crosses zero
at $b=0.2$ fm at $\sqrt{s}=7$ TeV. This is consistent with the result \cite{girom} of the Tevatron data analysis, in particular. The Pumplin bound can be rewritten in terms of $S(s,b)$  in the form
\begin{equation}\label{pbsf}
 \sigma_{diff}(s,b)\leq \frac{1}{4}S(s,b)(1-S(s,b)). 
 \end{equation}
This inequality clearly indicates that the Pumplin bound cannot be applied in the region where $S(s,b)$ is negative. It should be noted here
that this region is determined by the interval $0<b<R(s)$, where $R(s)$ is the solution of the equation $S(s,b)=0$. In this impact parameter
range only trivial bound 
\[
 \sigma_{diff}(s,b)\leq \sigma_{inel}(s,b)
\]
can be applied.
But, at $b\geq R(s)$ the scattering is absorptive and therefore the original Pumplin bound should be valid. The integrated bound
will be modified, however. Namely, in this case it should be written in the form
\begin{equation}\label{pumprefl}
\bar \sigma_{diff}(s)\leq \frac{1}{2}\bar \sigma_{tot}(s)-\bar \sigma_{el}(s),
 \end{equation}
where  $\bar \sigma_i(s)$ are the reduced cross-sections:
 \[
 \bar \sigma_i(s)\equiv \sigma_i(s) -8\pi\int_0^{R(s)}bdb\sigma_i(s,b),
 \]
 and $i\equiv diff, tot, el$, respectively.

Thus, there is no  inconsistency between the saturation of unitarity limit leading to Eq. (\ref{rd}) and the Pumplin bound for the inelastic
diffraction cross--section.

\end{document}